\pdfoutput=1
\newcommand*{\ATLASLATEXPATH}{}
\documentclass[cernpreprint,texlive=2011,txfonts,UKenglish]{atlasdoc}

\usepackage[subfigure=true]{\ATLASLATEXPATH atlaspackage}
\usepackage{\ATLASLATEXPATH atlasbiblatex}

\usepackage{\ATLASLATEXPATH atlascontribute}

\usepackage{\ATLASLATEXPATH atlasphysics}

\usepackage{hepnames}
\usepackage{mathtools}
\usepackage{multirow}
\renewcommand{\arraystretch}{1.15}

\usepackage{ZZEarlyPAPER-defs}

\hypersetup{pdftitle={ATLAS draft},pdfauthor={The ATLAS Collaboration}}

\AtlasTitle{Measurement of the \ZZ{} Production Cross Section in $\Pproton\Pproton$ Collisions at $\sqrt{s} = 13$~\TeV{} with the ATLAS Detector}

\author{The ATLAS Collaboration}

\AtlasRefCode{STDM-2015-13}

\AtlasNote{ATL-COM-PHYS-2015-1313}

\PreprintIdNumber{CERN-PH-EP-2015-318}

\AtlasJournal{Phys.~Rev.~Lett.}

\AtlasAbstract{%
The \ZZ{} production cross section in proton--proton collisions at 13~\TeV{} center-of-mass energy is measured using 3.2~\ifb of data recorded with the ATLAS detector at the Large Hadron Collider. The considered \PZ{} boson candidates decay to an electron or muon pair of mass 66--116~\GeV{}.
The cross section is measured in a fiducial phase space reflecting the detector acceptance. It is also extrapolated to a total phase space for \PZ{} bosons in the same mass range and
of all decay modes, giving $16.7^{+2.2}_{-2.0}\text{(stat.)}^{+0.9}_{-0.7}\text{(syst.)}^{+1.0}_{-0.7}\text{(lumi.)}$~pb. The results agree with standard model predictions.
}

\AtlasCoverSupportingNote{Measurement of the \ZZ{} Production Cross Section in $\Pproton\Pproton$ Collisions at $\sqrt{s} = 13$~\TeV{} with the ATLAS Detector}{https://cds.cern.ch/record/2055805}
\AtlasCoverCommentsDeadline{8th December 2015}

\AtlasCoverAnalysisTeam{Maurice Becker, Andrea Bocci, Will Buttinger,
Tiesheng Dai, Denys Denysiuk, Roberto Di Nardo, Cong Geng, Samira
Hassani, Naoki Kimura, Kostas Kordas, Alexandros Marantis, Emily
Nurse, Marc-Andre Pleier, Stefan Richter, Jonatan Rosten, Anton Wolf,
Yusheng Wu, Bing Zhou}

\AtlasCoverEdBoardMember{Kevin Black~(chair), Walter Hopkins, Laurent Serin}

\AtlasCoverEgroupEditors{atlas-stdm-2015-13-editors@cern.ch}

\AtlasCoverEgroupEdBoard{atlas-stdm-2015-13-editorial-board@cern.ch}

\begin{document}

\maketitle



Studying the production of pairs of \PZ{} bosons in proton--proton ($\Pproton\Pproton$)
interactions at the Large Hadron Collider (LHC)
tests the electroweak sector of the standard model
(SM) at the highest available energies.
In $\Pproton\Pproton$ collisions at a center-of-mass energy of
$\sqs=13$~\TeV{}, \ZZ{} production is dominated by quark--antiquark ($\Pquark\APquark$)
interactions, with an $\mathcal{O}(10\%)$ contribution from loop-induced
gluon--gluon ($\Pgluon\Pgluon$)
interactions~\cite{Grazzini:2015hta,Caola:2015psa}.
The SM \ZZ{} production can proceed via a Higgs boson propagator,
although this contribution is suppressed in the region where both \PZ{}
bosons are produced on-shell. 
As such, non-Higgs \ZZ{} production is an important background in
studies of the Higgs boson~\cite{Aad:2015lha,Aad:2015xua,Khachatryan:2015mma}. It is also a background in searches for new physics
producing pairs of \PZ{} bosons at high invariant mass~\cite{Aad:2015kna,Khachatryan:2014gha} and sensitive to triple neutral-gauge-boson couplings, which are not allowed in the SM~\cite{Baur:2000ae}.


This Letter presents the first measurement of the $\ZZ$ production cross section
in $\Pproton\Pproton$ interactions at 
$\sqs=13$~\TeV{}. Throughout it, ``\PZ{} boson'' refers to the superposition of a \PZ boson and virtual photon with mass in the range 66--116~\GeV{}.
The analyzed data correspond to an integrated luminosity of $3.2 \pm
  0.2$~\ifb{}, 
  collected with the ATLAS detector~\cite{PERF-2007-01}. 
 The uncertainty of the integrated luminosity is derived, following a methodology similar to that detailed in Ref.~\cite{Aad:2013ucp}, from a preliminary calibration of the luminosity scale using a pair of $x$--$y$ beam-separation scans performed in June 2015.
The \ZZ{} production cross section was previously measured at
$\sqs=7$ and 8~\TeV{} by the ATLAS and CMS
collaborations~\cite{Aad:2012awa,CMS:2014xja,Chatrchyan:2012sga}
and found to be consistent with SM predictions. 


Candidate events are reconstructed in the
fully leptonic $\ZZllll$ decay channel where
$\ell$ and $\ell'$ can be an electron or a muon.
The cross section \sigmafid{} is found by counting candidate events,
subtracting the expected contribution from background events,
correcting for detector effects, 
and dividing by
the integrated luminosity. 
It is measured in a fiducial phase space that
corresponds closely to the experimental acceptance.
In addition, an
extrapolation of the cross section to a total phase space for \PZ bosons, 
\sigmatot{}, is performed.
The presented cross-section measurements are inclusive with respect to
additional jets. 
Small contributions from triboson production with two
leptonically decaying \Z{} bosons and a third hadronically decaying
weak boson and contributions from double parton
scattering  are included in the measurement.

The fiducial phase space, which is designed to reflect the acceptance of the
ATLAS detector (described below), is 
defined for simulated 
events by applying the following criteria
to the final-state particle-level objects.
Final-state electrons and muons are required to be prompt
(i.e.~to not originate from hadron or $\tau$ decay) and their kinematics are computed
including the contributions from prompt photons with a distance
in $\eta$--$\phi$ coordinates\footnote{ATLAS uses a right-handed coordinate
  system with its origin at the nominal interaction point in the
  center of the detector and the $z$-axis along the beam pipe. The
  $x$-axis points to the center of the LHC ring, and the $y$-axis
  points upward. Cylindrical coordinates $(r,\phi)$ are used in the
  transverse plane, $\phi$ being the azimuthal angle around the $z$-axis.
  The pseudorapidity is defined in terms of the polar angle
  $\theta$ as $\eta=-\ln[\tan(\theta/2)]$.}
of $\Delta R_{\ell,\gamma} = \sqrt{(\Delta\eta_{\ell,\gamma})^2 + (\Delta\phi_{\ell,\gamma})^2}< 0.1$
between the charged lepton and the photon, as motivated in
Ref.~\cite{ATL-PHYS-PUB-2015-013}. The leptons are required to be well-separated
with $\Delta R_{\ell,\ell'}>0.2$ between any two leptons. 
Each lepton must have a momentum component transverse to the beam
direction
 $\pta > 20$~\GeV{} and pseudorapidity $|\eta| < 2.7$.
Events must have exactly four leptons satisfying the above criteria forming two pairs of same-flavor oppositely charged
leptons ($\mumu$ or $e^{+}e^{-}$). 
This gives rise to three signal channels: \eeee{}, \mmmm{}, and \eemm{}.
Each lepton pair must have an invariant mass in the range 66--116~\GeV{}. In the \eeee{} and \mmmm{} channels, where there are two possible ways to form same-flavor oppositely charged lepton pairs, the combination that minimizes $|\mlla - \mz| + |\mllb - \mz|$ is chosen, where \mlla{} and \mllb{}  are the invariant masses
of the lepton pairs and \mz{} is the mass of the \PZ{} boson. 

The ATLAS detector is a multipurpose particle detector with a
cylindrical geometry. It consists of
layers of inner tracking detectors, calorimeters, and muon chambers. 
The inner detector (ID) 
covers the
pseudorapidity range $|\eta| < 2.5$. The calorimeter covers the
pseudorapidity range $|\eta| < 4.9$. Within $|\eta| <2.47$ the finely segmented electromagnetic calorimeter identifies electromagnetic showers and measures
their energy and position, providing electron identification together with the ID. 
The muon spectrometer (MS) surrounds the calorimeters and provides muon identification and
measurement in the region $|\eta| < 2.7$ and triggering in the region  $|\eta| < 2.4$. 

A muon is reconstructed by matching a track (or track segment)
reconstructed in the MS to
a track reconstructed in the ID.  
Its momentum is calculated
by combining the information from the two systems and correcting for energy deposited in the
calorimeters. 
In regions of limited coverage of the MS ($|\eta| < 0.1$) or outside
the ID acceptance 
($2.5 < |\eta| < 2.7$), muons can also be reconstructed by matching
calorimeter signals consistent with muons to ID tracks
(calorimeter-tagged muons) or standalone in the
MS~\cite{cite:muon-perf-pub}, respectively.


An electron is reconstructed from an energy deposit (cluster) in the
electromagnetic calorimeter matched to a track in the ID. Its momentum
is computed from the cluster  energy and
the direction of the track.
Electrons are distinguished from other particles using several
identification criteria that rely on the shapes of electromagnetic
showers as well as tracking and track-to-cluster matching
quantities. The output of a likelihood function taking these
quantities as input, similar to that
described in  Ref.~\cite{cite:electron}, is used to identify electrons. 
Electrons sharing an ID track with a selected muon are ignored. 

The leptons are required to be isolated from other
particles using ID track information, and for muons also calorimeter
information (since standalone muons are outside the ID
acceptance). The exact requirements depend on the lepton \pta{} and
$\eta$ and are designed to give a uniform 99\% efficiency.

Leptons are required to originate from the primary vertex, defined as
the reconstructed vertex with the largest sum of the $\pta^{2}$ of the
associated tracks. To this end, the longitudinal impact parameter of
each lepton track, calculated with respect to the vertex
and multiplied by $\sin\theta$ of the track, is required to
be less than 0.5~mm. Furthermore, the significance
of the transverse impact parameter calculated with respect to the beam
line is required to be less than three (five) for muons (electrons).
Standalone muons are exempt from both impact parameter
requirements, as they do not have an ID track.

Candidate events are preselected by either a single-muon or dielectron trigger.
As in the fiducial phase space described above, leptons must have $\pta > 20$~\GeV{}.
There are slight differences from the fiducial phase space: electrons must satisfy $|\eta| <
2.47$ due to the limited experimental acceptance, and at least one
muon in the \mmmm{} channel must satisfy $|\eta| < 2.4$, corresponding
to the acceptance of the muon trigger. 
The other muons must satisfy $|\eta| < 2.7$.
Events are ignored if more than one selected muon is calorimeter-tagged or standalone.
Apart from the above differences, reconstructed candidate events are selected using exactly the same criteria that define the fiducial phase space.
A total of 63 events are observed, of which 15, 30, and 18 are in the \eeee{}, \eemm{}, and \mmmm{} channel, respectively. 



Monte-Carlo-simulated (MC) event samples are used  to obtain corrections
for detector effects and to estimate background contributions.
The principal signal sample is generated with the \POWHEG{} method and
framework~\cite{Nason:2004rx,Frixione:2007vw,Alioli:2010xd}, with a
diboson event generator \cite{Melia:2011tj,Nason:2013ydw} used to
simulate the $\PZ\PZ$ production process at
 next-to-leading order (NLO).\footnote{Throughout this Letter, orders of calculations refer to perturbative expansions in the strong coupling constant \alphas{} unless stated otherwise.}
The simulation of parton showering, of the underlying event, and of hadronization
is performed with \py{} \cite{Sjostrand:2006za,Sjostrand:2007gs} 
using the \aznlo{} set of tuned parameters (tune)~\cite{Aad:2014xaa}. 
\SHERPA{}~\cite{Gleisberg:2008ta,Hoeche:2009rj,Gleisberg:2008fv,Schumann:2007mg,Schonherr:2008av,Cascioli:2011va,Hoeche:2012yf} is used to generate a sample with the $\Pquark\APquark$-initiated process simulated at
NLO for \ZZ{} plus zero or one additional jet 
and at leading order (LO) for two or three additional jets, as well as a  
sample with the
loop-induced \gluongluon-initiated process simulated at
LO
with zero or one additional jet. 
These are used to include the loop-induced \gluongluon-initiated
production, which is not included in the \POWHEGpy{} sample, as well as to estimate, by comparison of the various samples, a systematic uncertainty due to the choice of event
generator.
The \cten{} NLO \cite{Lai:2010vv} parton distribution functions (PDFs) are used
in the event generation for all samples  above.
Additional samples are generated to estimate the contribution from
background events.
Triboson events are simulated with \SHERPA{}, using \cten{} PDFs, and 
$\Ptop\APtop\kern0.5pt\PZ$ events are simulated with \MADGRAPH~\cite{Alwall:2011uj} interfaced with
\py{} using the \nnpdf{} PDFs~\cite{Ball:2012cx} and the \afourteen{} tune~\cite{cite:autwo}.

In all MC samples, additional $\Pproton\Pproton$ interactions occurring in the same bunch crossing as the
\ZZ{} production, or in nearby ones, are simulated with \py{} with
\mstwlo{} PDFs~\cite{Martin:2009iq} and the \atwo{} tune~\cite{ATL-PHYS-PUB-2012-003}. 
The samples are then passed through a simulation of the ATLAS detector~\cite{SOFT-2010-01}
based on \GEANT{} 4~\cite{Agostinelli:2002hh}. 
Scale factors are applied to the simulated events to correct for the
small differences from data in the trigger, reconstruction,
identification, isolation, and impact parameter efficiencies for electrons and
muons~\cite{cite:muon-perf-pub, cite:electron}. Furthermore, the lepton momentum scales and
resolutions are adjusted to match the data.

Background events from processes with at least four prompt leptons in
the final state are estimated
with the MC samples described above, including uncertainties from the  cross-section
values, luminosity, and reconstruction effects.
Contributions of $0.07 \pm 0.02$ events from \ZZ{} processes where at
least one \Z{} boson decays to $\tau$ leptons, $0.17 \pm 0.05$ events from non-hadronic triboson processes,
 and $0.30 \pm 0.09$ events from all-leptonic $\Ptop\APtop\PZ$ processes are predicted. 
Events from processes with two or three prompt leptons, e.g. \PZ{},
\WW{}, \WZ{}, $\Ptop\APtop$, and \ZZ{} events
where one \PZ{} boson decays hadronically, where associated jets or
photons contain or fake a nonprompt lepton, can pass the event
selection. 
This background contribution is estimated to be
$0.09^{+1.08}_{-0.04}$ events, using control
samples and a data-driven technique described in Ref.~\cite{Aad:2012awa}. 
The uncertainty is dominated by the small number of events in the control
samples. It can be asymmetric due to truncation, as background contributions cannot be negative.
Background from two single \PZ{} bosons produced in different $\Pproton\Pproton$ collisions in the same bunch crossing is estimated to be negligible.
The total expected number of background events is  $0.20 \pm 0.05$
($0.25^{+0.40}_{-0.05}$, $0.17^{+1.00}_{-0.04}$) 
in the \eeee{} (\eemm{}, \mmmm{}) channel, giving a total of
$0.62^{+1.08}_{-0.11}$ events.

A factor \czz{} is applied to correct for detector inefficiencies
and resolution effects. It relates the background-subtracted number of selected events to the
number in the fiducial phase space, and is defined as the ratio of generated signal events passing the selection criteria using
reconstructed objects to the
number passing the fiducial criteria using generator-level objects. \czz{} is determined with a combination of the \POWHEG{} \ZZ{} MC sample
and  the \sherpa{} loop-induced \gluongluon-initiated 
sample. The normalization of the latter is scaled to $\mathcal{O}(\alphas^{3})$ accuracy~\cite{Caola:2015psa} in order to improve the model used to correct the measurement. 
The \czz{} value and its total uncertainty is determined to be $0.55 \pm 0.02$
($0.63 \pm 0.02$, $0.81 \pm 0.03$) in the \eeee{} (\eemm{}, \mmmm{}) channel.
The dominant systematic uncertainties come from the uncertainties of
the scale factors used to correct lepton reconstruction and identification
efficiencies in the simulation and the choice of MC generator. Other smaller uncertainties come from the scale and resolution of the lepton
  momenta, PDFs, and statistical fluctuations in the MC sample.
Table~\ref{tab:czz_uncert_overview} gives a breakdown of the systematic uncertainties.
\begin{table}[h!]
\centering
	\caption{Relative uncertainties of the correction factor \czz{} by signal channel, expressed in percent.}
\begin{tabular}{llll}
         \toprule
Source	& \eeee{}	& \eemm{}	& \mmmm{}\\\midrule
Statistical (signal samples)	& $0.7$	& $0.5$	& $0.5$\\
Theoretical (generator, PDFs)	& $2.5$	& $2.5$	& $2.5$\\
Experimental efficiencies		& $2.3$	& $2.2$	& $2.0$\\
Momentum scales and resolutions	& $0.4$	& $0.2$	& $0.1$\\
\midrule
Total	& $3.5$	& $3.3$	& $3.2$\\
         \bottomrule
	\end{tabular}
        \label{tab:czz_uncert_overview}
\end{table}


Figure~\ref{fig:ZZproperties} shows the invariant mass
of the leading-\ptll{} and the subleading-\ptll{} lepton pair ($\ell\ell$),  as well as the
invariant mass, transverse momentum, and rapidity of the four-lepton
system.
Distributions from data are compared to the signal and background
expectations, with good agreement in general.

\begin{figure}
	\begin{center}
	\subfigure[]{
          \includegraphics[width=0.48\textwidth]{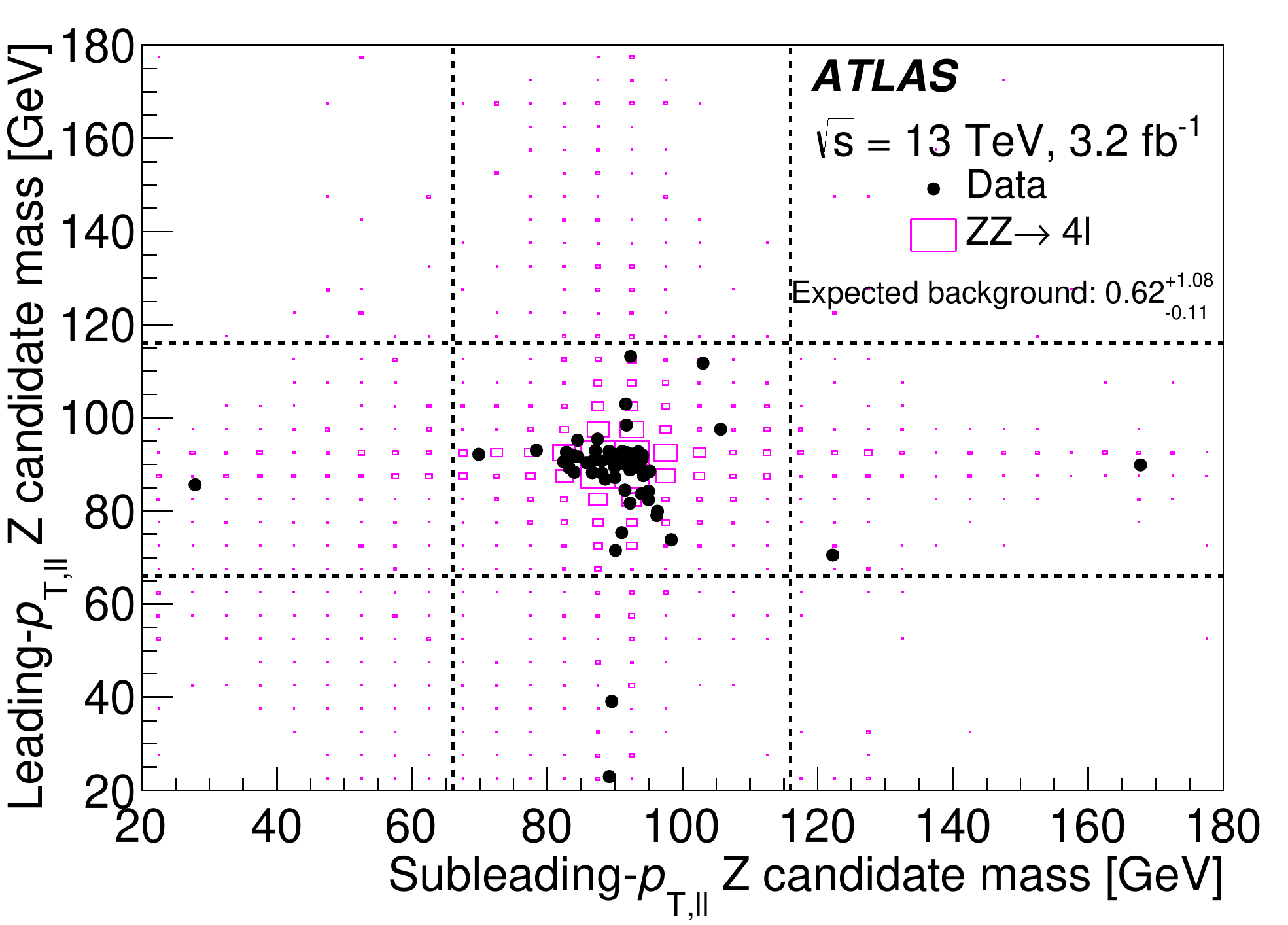}
                \label{fig:2dZZ}
	}
	\subfigure[]{
         \includegraphics[width=0.48\textwidth]{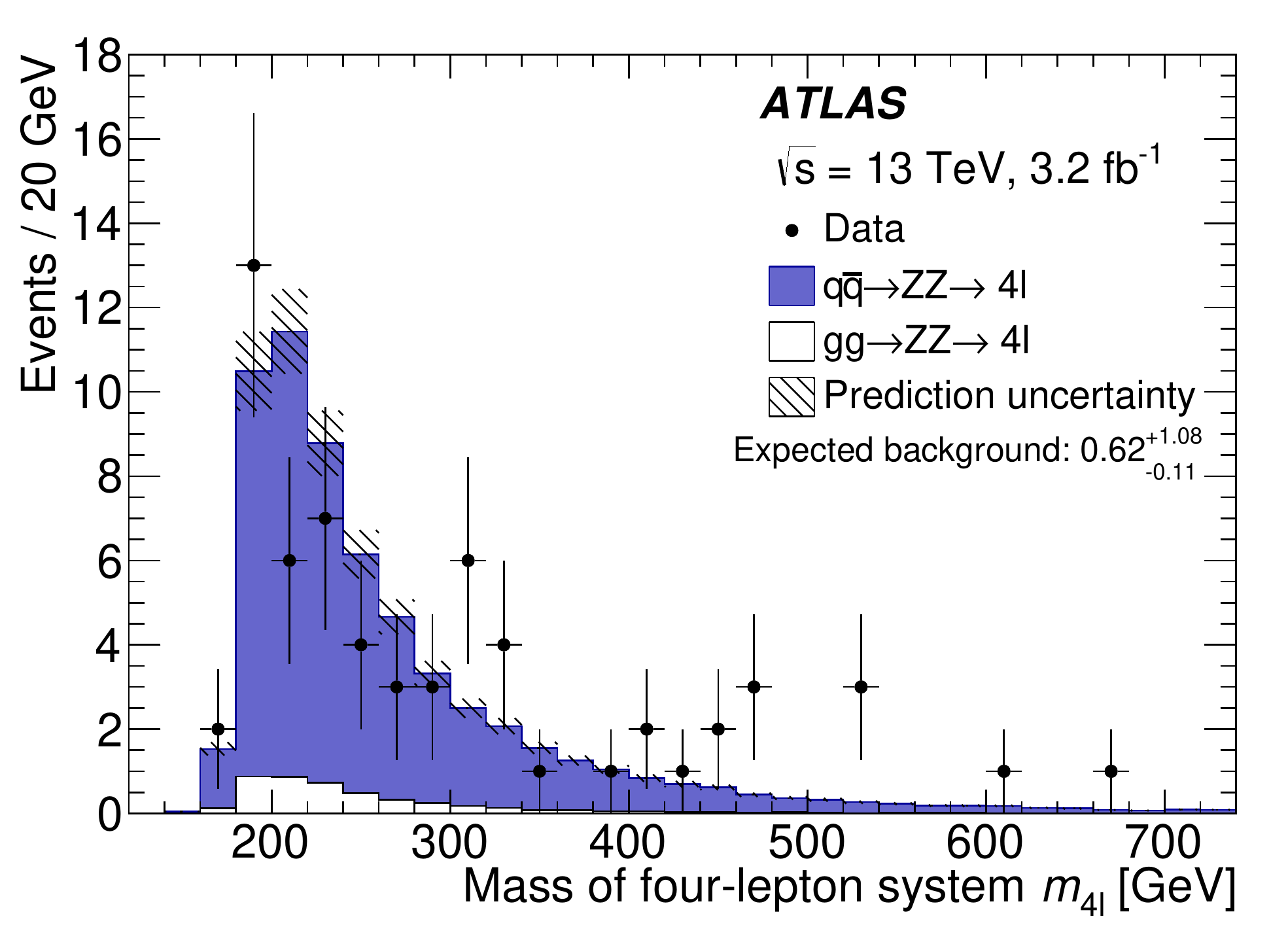}
                \label{fig:ZZMass}
	}
	\subfigure[]{
                \includegraphics[width=0.48\textwidth]{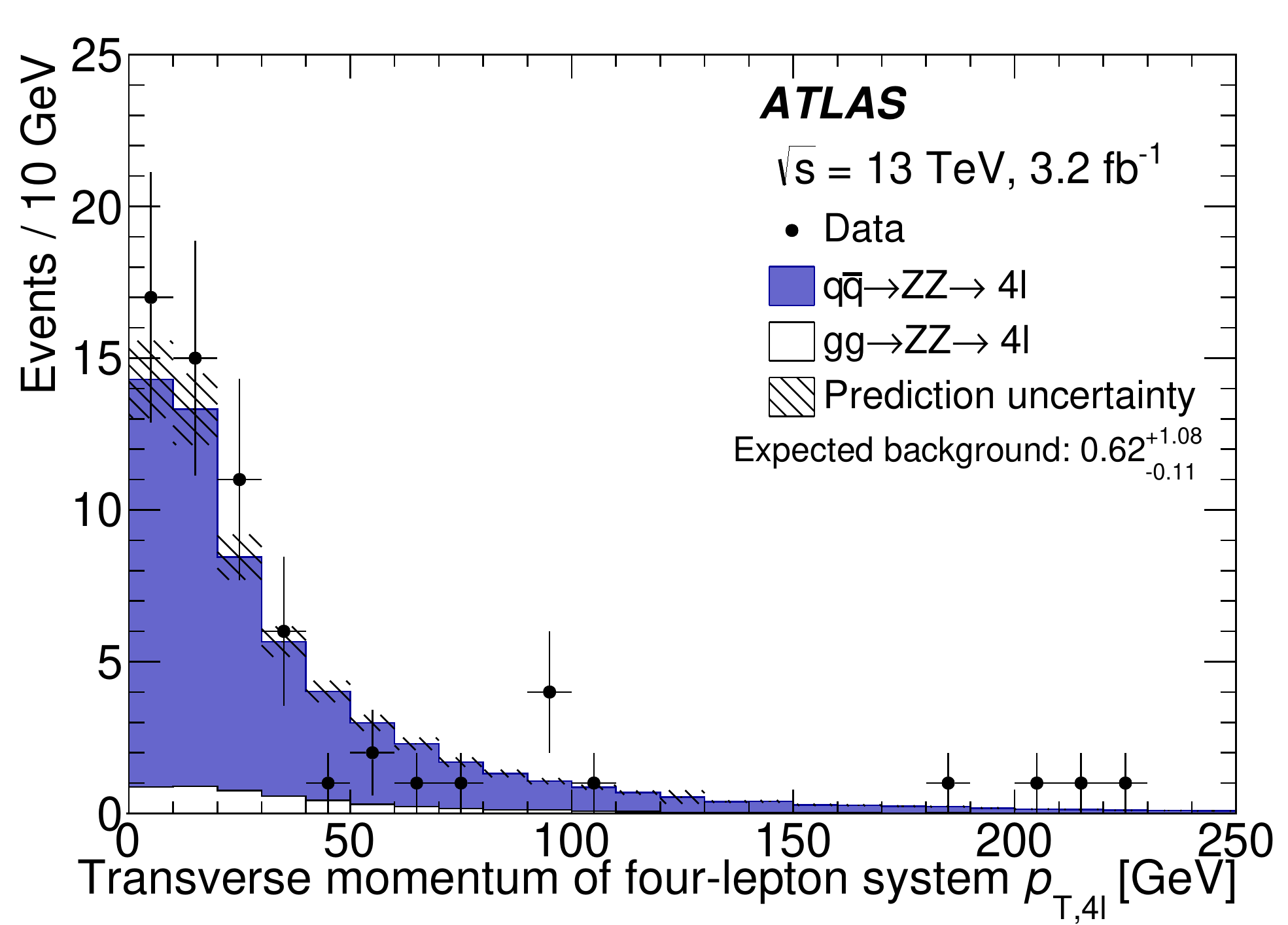}
                \label{fig:ZZpT}
	}
	\subfigure[]{
               \includegraphics[width=0.48\textwidth]{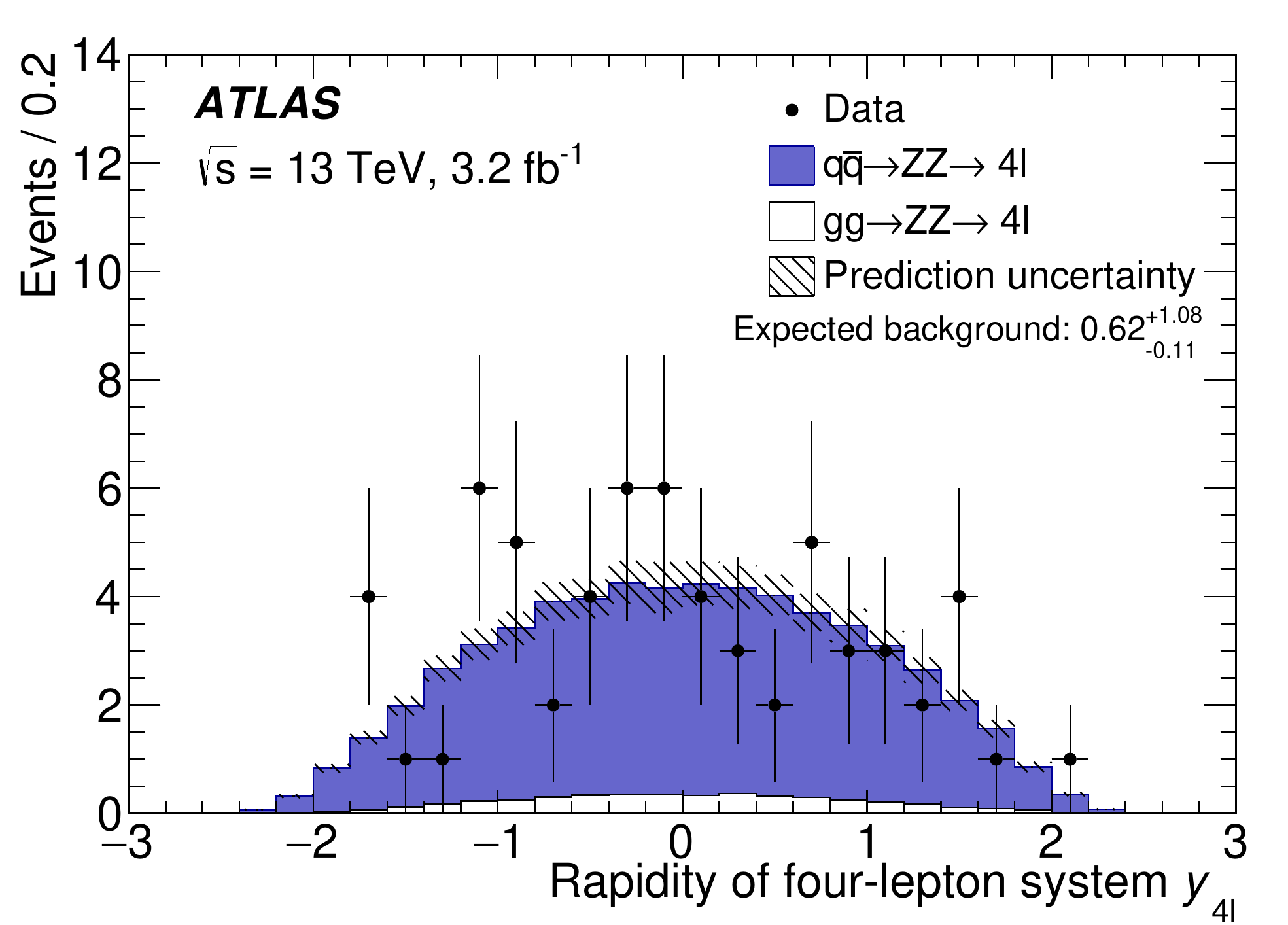}
                \label{fig:ZZy}
	}
	\end{center}
	\caption {(a) Invariant mass $\mll$ of the leading-\ptll{}
          versus the
           subleading-\ptll{} lepton pair ($\ell\ell$), before the requirement $66~\text{\GeV{}} < \mll < 116~\text{\GeV{}}$ is
           applied. The dashed lines indicate this requirement. (b) Invariant mass, (c) transverse momentum, and (d)
          rapidity of the four-lepton system in selected
          events. The points represent experimental data. The filled
          histograms show the signal prediction from simulation, including the $\Pquark\APquark$ and loop-induced
          $\Pgluon\Pgluon$-initiated process. The contributions are stacked. 
In the simulation, the prediction from \POWHEGpy{} combined with
\SHERPA{} is scaled to the $\mathcal{O}(\alphas^{2})$ prediction.
The
uncertainties in the simulation are from the same sources as the
\czz{} uncertainty.
In addition, 6\% $\PZ\PZ$ cross-section uncertainty and 5\%{} integrated-luminosity uncertainty are included. The expected background of $0.62^{+1.08}_{-0.11}$ events is not shown as a histogram due to its small size.}
  \label{fig:ZZproperties}
\end{figure}

The fiducial cross section is determined using a maximum-likelihood
fit to the event counts in the three signal channels.
A Poisson probability function is used to parametrize the number of expected events, multiplied by Gaussian distributions that model the nuisance parameters representing systematic uncertainties. 
This procedure can lead to asymmetric uncertainties as Poisson-distributed
variables cannot be negative.

The cross section measured in the fiducial phase space is also
extrapolated to the total phase space, which includes a correction for QED
final-state radiation effects.
The extrapolation factor is obtained from the same combination of MC
samples as used in the \czz{} determination.
The ratio of the fiducial to full phase-space cross section is
$0.39 \pm 0.02$, in all three channels. It is corrected for the
$\sim$3\% increase bias introduced by the pairing algorithm in the \eeee{} and \mmmm{} channels.
The dominant systematic uncertainty comes from the difference between the nominal value
and that obtained using the \sherpa{} samples. Smaller uncertainties
are derived from PDF variations in the \cten{} error set, differences between using \py{} and
\HERWIGpp{}~\cite{Gieseke:2011na} for simulating the rest of the
event, and varying the QCD renormalization and factorization scales independently by a factor of two.
In order to
extrapolate to the total cross section, the fiducial cross sections
are divided by the ratio $0.39 \pm 0.02$ and corrected for the leptonic branching
fraction\footnote{This value excludes $\gamma^*$ contributions. Including these, the branching fraction $\ZZllll$ is about 1.01--1.02 times larger.} $(3.3658\%)^{2}$~\cite{Agashe:2014kda}.

The measured fiducial cross sections are shown in Table~\ref{tab:results} and
Figure~\ref{fig:theoryComparison} along with a comparison to
$\mathcal{O}(\alphas^{2})$
calculations~\cite{Grazzini:2015hta}. Table~\ref{tab:results} also shows the total combined cross
section.
The \cten{} next-to-next-to-leading order PDFs~\cite{Gao:2013xoa} and a dynamic scale equal to the mass of the four-lepton system are used in the calculation. 
The loop-induced \gluongluon-initiated process is included, and contributes
7.0\%\ (5.8\%) of the cross section in the fiducial (total) phase
space.
The predicted cross sections in the fiducial phase space are corrected for QED final-state radiation effects, which amount
  to a 4\% reduction.
The measurements agree with the SM predictions.


The theoretical predictions do not include the following effects.
The loop-induced \gluongluon-initiated process calculated at $\mathcal{O}(\alphas^{2})$ could
receive large corrections at $\mathcal{O}(\alphas^{3})$ of 70\%~\cite{Caola:2015psa}, which
would increase the prediction by 4--5\%. Electroweak corrections at
next-to-leading order~\cite{Bierweiler:2013dja,Biedermann:2016yvs} are expected to reduce the cross section by
7--8\%~\cite{Biedermann:2016yvs}. Furthermore, the contribution from double parton
scattering is
not accounted for, but is expected to be an effect of less than 1\%~\cite{Aad:2013bjm}.


The measured total cross section is compared to measurements at lower
center-of-mass energies and to a prediction from
\MCFM~\cite{Campbell:1999ah} with the \cfourteen{} NLO PDFs~\cite{Dulat:2015mca},
which is calculated at $\mathcal{O}(\alphas^{1})$
accuracy for the $\Pquark\APquark$-initiated process
     and at $\mathcal{O}(\alphas^{2})$ accuracy for the loop-induced \gluongluon-initiated process
      and is shown
versus center-of-mass energy in Figure~\ref{fig:exptComparison}.
The cross section increases by a factor of more than two with a center-of-mass energy increase from 8~\TeV{} to 13~\TeV{}.

\renewcommand{\arraystretch}{1.5}
\begin{table}[h!]
\begin{center}
	\caption{Cross-section measurement results compared to the
          $\mathcal{O}(\alphas^{2})$ standard model predictions. The per-channel and combined
        fiducial cross sections are shown along with the combined
        total cross section. 
      For experimental results, the statistical, systematic, and luminosity
      uncertainties are shown. 
      For theoretical predictions, the PDF and
      renormalization and factorization scale uncertainties
      added in quadrature are shown.}
	\begin{tabular}{lrr} 
         \toprule
         \multicolumn{2}{c}{Measurement} & \multicolumn{1}{c}{$\mathcal{O}(\alphas^{2})$ prediction} \\
         \midrule  
         $\sigmafideeee$                &  8.4     $^{+2.4}_{-2.0} $(stat.) $^{+0.4}_{-0.2} $(syst.) $^{+0.5}_{-0.3} $(lumi.)  fb &  $6.9^{+0.2}_{-0.2}$ fb\\
         $\sigmafideemm$                & 14.7     $^{+2.9}_{-2.5} $(stat.) $^{+0.6}_{-0.4} $(syst.) $^{+0.9}_{-0.6} $(lumi.)  fb & $13.6^{+0.4}_{-0.4}$ fb\\
         $\sigmafidmmmm$                &  6.8     $^{+1.8}_{-1.5} $(stat.) $^{+0.3}_{-0.3} $(syst.) $^{+0.4}_{-0.3} $(lumi.)  fb &  $6.9^{+0.2}_{-0.2}$ fb\\
         $\sigmafid$                    & 29.7     $^{+3.9}_{-3.6} $(stat.) $^{+1.0}_{-0.8} $(syst.) $^{+1.7}_{-1.3} $(lumi.)  fb & $27.4^{+0.9}_{-0.8}$ fb\\
         $\sigmatot$                    & 16.7     $^{+2.2}_{-2.0} $(stat.) $^{+0.9}_{-0.7} $(syst.) $^{+1.0}_{-0.7} $(lumi.)  pb & $15.6^{+0.4}_{-0.4}$ pb\\
          \bottomrule 
	 \end{tabular}
        \label{tab:results}
\end{center}
\end{table}

\begin{figure}[h]
	\centering
          \subfigure[]{
          \includegraphics[width=0.495\textwidth]{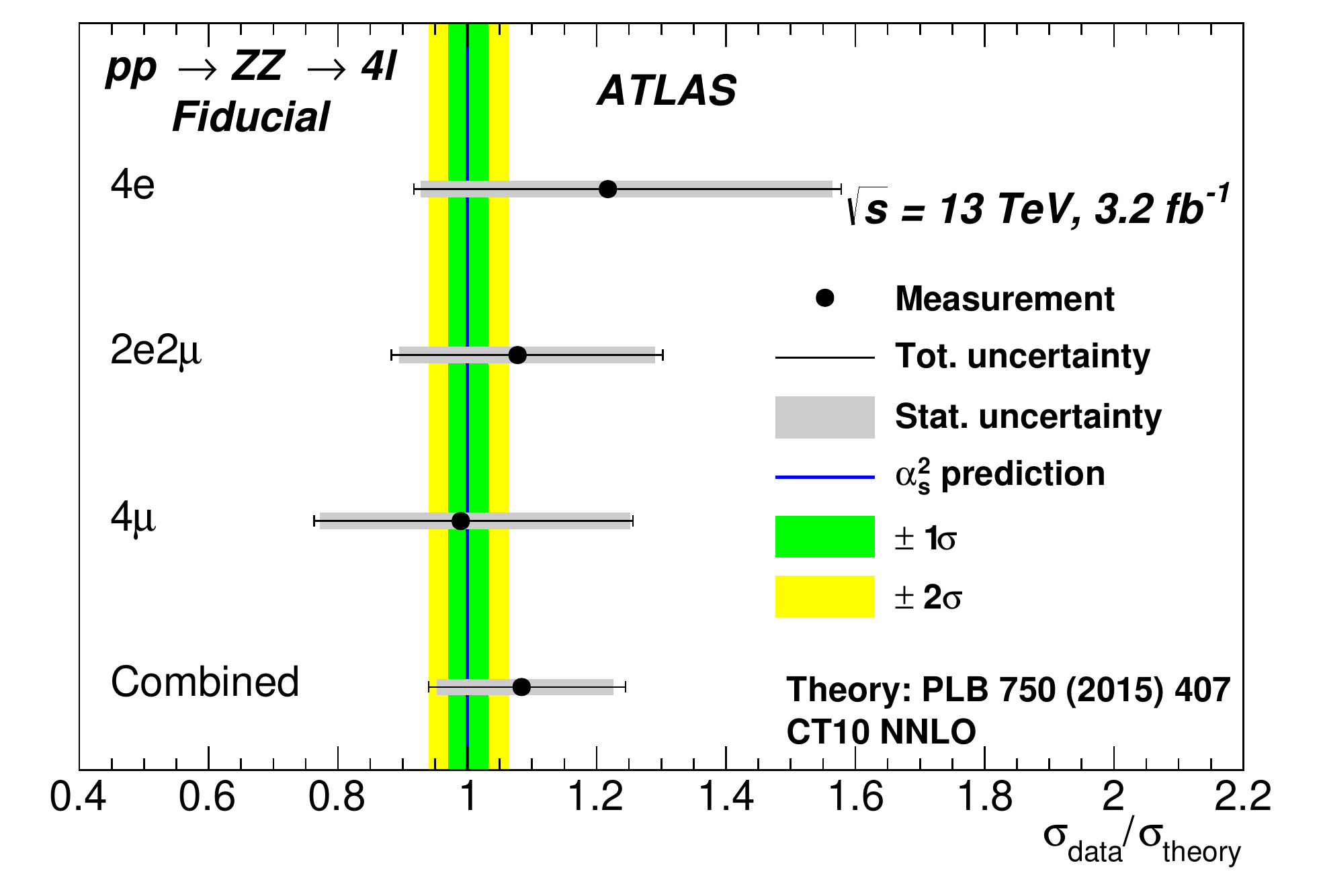}
          \label{fig:theoryComparison}
        }
        \subfigure[]{
         \includegraphics[width=0.468\textwidth]{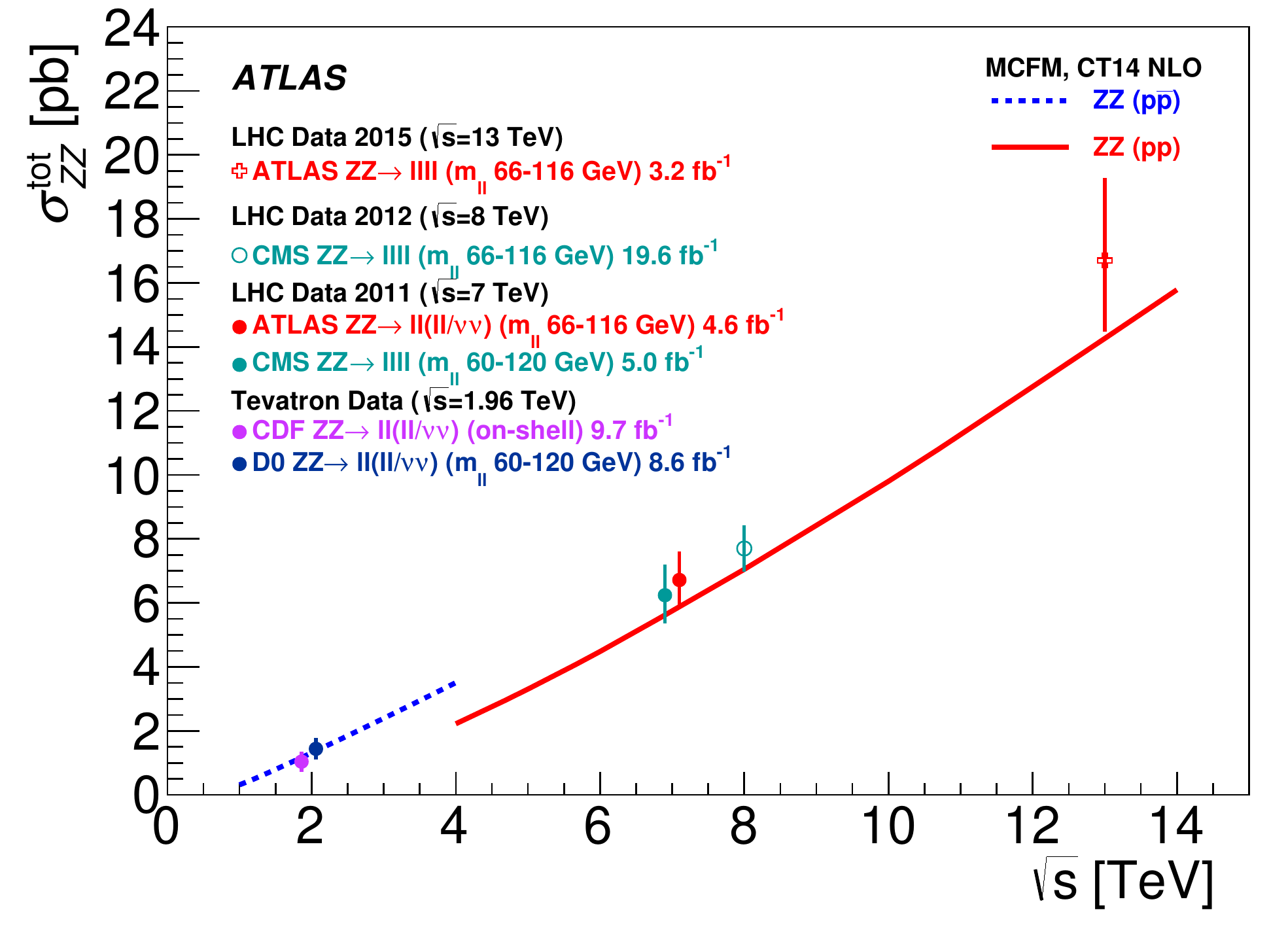}
          \label{fig:exptComparison}
		}
    \caption{(a) Comparison between measured fiducial cross sections and $\mathcal{O}(\alphas^{2})$
      predictions. (b) Total cross section compared to measurements at lower center-of-mass energies by ATLAS, CMS, CDF,
      and D0~\cite{Aad:2012awa,CMS:2014xja,Chatrchyan:2012sga,Aaltonen:2014yfa,Abazov:2012cj}, and to a prediction from \MCFM{} at $\mathcal{O}(\alphas^{1})$ accuracy for the $\Pquark\APquark$-initiated process and at $\mathcal{O}(\alphas^{2})$ accuracy for the loop-induced \gluongluon-initiated process. A full $\mathcal{O}(\alphas^{2})$ prediction (known to improve agreement at $\sqs = 13$~\TeV{}) was not yet available for all the different center-of-mass energies. Some data points are shifted horizontally to improve readability. The $\PZ\PZ$ cross section as function of $\sqrt{s}$ in the range 130--209~\GeV{} was also measured at the LEP 2 $\Ppositron\Pelectron$ collider \cite{Schael:2013ita}.}
\label{fig:Comparison}
\end{figure}

In summary, ATLAS has measured the \ZZ{} production cross section in $3.2$~\ifb{} of 13~\TeV{} $\Pproton\Pproton$ collisions at the LHC using the fully leptonic
decay channel $\ZZllll$. Fiducial cross sections as
well as a total cross section for \Z{} bosons with mass 66--116~\GeV{} have been
measured and agree well with $\mathcal{O}(\alphas^{2})$ SM predictions.



We thank CERN for the very successful operation of the LHC, as well as the
support staff from our institutions without whom ATLAS could not be
operated efficiently.

We acknowledge the support of ANPCyT, Argentina; YerPhI, Armenia; ARC, Australia; BMWFW and FWF, Austria; ANAS, Azerbaijan; SSTC, Belarus; CNPq and FAPESP, Brazil; NSERC, NRC and CFI, Canada; CERN; CONICYT, Chile; CAS, MOST and NSFC, China; COLCIENCIAS, Colombia; MSMT CR, MPO CR and VSC CR, Czech Republic; DNRF, DNSRC and Lundbeck Foundation, Denmark; IN2P3-CNRS, CEA-DSM/IRFU, France; GNSF, Georgia; BMBF, HGF, and MPG, Germany; GSRT, Greece; RGC, Hong Kong SAR, China; ISF, I-CORE and Benoziyo Center, Israel; INFN, Italy; MEXT and JSPS, Japan; CNRST, Morocco; FOM and NWO, Netherlands; RCN, Norway; MNiSW and NCN, Poland; FCT, Portugal; MNE/IFA, Romania; MES of Russia and NRC KI, Russian Federation; JINR; MESTD, Serbia; MSSR, Slovakia; ARRS and MIZ\v{S}, Slovenia; DST/NRF, South Africa; MINECO, Spain; SRC and Wallenberg Foundation, Sweden; SERI, SNSF and Cantons of Bern and Geneva, Switzerland; MOST, Taiwan; TAEK, Turkey; STFC, United Kingdom; DOE and NSF, United States of America. In addition, individual groups and members have received support from BCKDF, the Canada Council, CANARIE, CRC, Compute Canada, FQRNT, and the Ontario Innovation Trust, Canada; EPLANET, ERC, FP7, Horizon 2020 and Marie Sk\l{}odowska-Curie Actions, European Union; Investissements d'Avenir Labex and Idex, ANR, Region Auvergne and Fondation Partager le Savoir, France; DFG and AvH Foundation, Germany; Herakleitos, Thales and Aristeia programmes co-financed by EU-ESF and the Greek NSRF; BSF, GIF and Minerva, Israel; BRF, Norway; the Royal Society and Leverhulme Trust, United Kingdom.

The crucial computing support from all WLCG partners is acknowledged
gratefully, in particular from CERN and the ATLAS Tier-1 facilities at
TRIUMF (Canada), NDGF (Denmark, Norway, Sweden), CC-IN2P3 (France),
KIT/GridKA (Germany), INFN-CNAF (Italy), NL-T1 (Netherlands), PIC (Spain),
ASGC (Taiwan), RAL (UK) and BNL (USA) and in the Tier-2 facilities
worldwide.

\clearpage


\printbibliography

\newpage
\begin{flushleft}
{\Large The ATLAS Collaboration}

\bigskip

G.~Aad$^\textrm{\scriptsize 85}$,
B.~Abbott$^\textrm{\scriptsize 112}$,
J.~Abdallah$^\textrm{\scriptsize 150}$,
O.~Abdinov$^\textrm{\scriptsize 11}$,
B.~Abeloos$^\textrm{\scriptsize 116}$,
R.~Aben$^\textrm{\scriptsize 106}$,
M.~Abolins$^\textrm{\scriptsize 90}$,
O.S.~AbouZeid$^\textrm{\scriptsize 136}$,
H.~Abramowicz$^\textrm{\scriptsize 152}$,
H.~Abreu$^\textrm{\scriptsize 151}$
% ...
\bigskip
\\
% ...
$^{11}$ Institute of Physics, Azerbaijan Academy of Sciences, Baku, Azerbaijan\\
% ...
$^{85}$ CPPM, Aix-Marseille Universit{\'e} and CNRS/IN2P3, Marseille, France\\
% ...
$^{90}$ Department of Physics and Astronomy, Michigan State University, East Lansing MI, United States of America\\
% ...
$^{106}$ Nikhef National Institute for Subatomic Physics and University of Amsterdam, Amsterdam, Netherlands\\
% ...
$^{112}$ Homer L. Dodge Department of Physics and Astronomy, University of Oklahoma, Norman OK, United States of America\\
% ...
$^{116}$ LAL, Univ. Paris-Sud, CNRS/IN2P3, Universit{\'e} Paris-Saclay, Orsay, France\\
% ...
$^{136}$ Santa Cruz Institute for Particle Physics, University of California Santa Cruz, Santa Cruz CA, United States of America\\
% ...
$^{150}$ Institute of Physics, Academia Sinica, Taipei, Taiwan\\
$^{151}$ Department of Physics, Technion: Israel Institute of Technology, Haifa, Israel\\
$^{152}$ Raymond and Beverly Sackler School of Physics and Astronomy, Tel Aviv University, Tel Aviv, Israel\\
% ...
\end{flushleft}


\end{document}